\documentclass[twocol]{ametsocV5}

\newcommand{\pr}{\mathbb{P}}
\newcommand{\ex}{\mathbb{E}}

\newcommand{\lcal}{\mathcal{L}}

\newcommand{\tr}{^\top}
\newcommand{\inv}{^{-1}}
\newcommand{\diag}{\text{diag}}

\newcommand{\wt}{\widetilde}
\newcommand{\ov}{\overline}
\newcommand{\re}{\text{Re}}
\newcommand{\im}{\text{Im}}

\newcommand{\pder}[2]{\frac{\partial{#1}}{\partial{#2}}}
\newcommand{\pdersq}[2]{\frac{\partial^2{#1}}{\partial{#2}^2}}

\newcommand{\bmt}{\begin{bmatrix}}
\newcommand{\emt}{\end{bmatrix}}

\newcommand{\st}{^*}

\newcommand{\R}{\mathbb{R}}

\definecolor{OliveGreen}{rgb}{0,0.6,0}
\definecolor{purple}{rgb}{0.8,0,0.8}

\newcommand{\aln}[1]{\begin{align*}#1\end{align*}}
\newcommand{\alnn}[1]{\begin{align}#1\end{align}}
\newcommand{\qp}{q^+}
\newcommand{\qm}{q^-}

\usepackage{amsmath,amsfonts,amssymb,bm}
\usepackage{mathptmx}
\usepackage{newtxtext}
\usepackage{newtxmath}

\title{Path properties of atmospheric transitions: illustration with a low-order
sudden stratospheric warming model}

\authors{Justin Finkel\correspondingauthor{Justin Finkel, jfinkel@uchicago.edu}}

\affiliation{Committee on Computational and Applied Mathematics, University of Chicago}

\extraauthor{Dorian S. Abbot}
\extraaffil{Department of the Geophysical Sciences, University of Chicago}
\extraauthor{Jonathan Weare}
\extraaffil{Courant Institute of Mathematical Sciences, New York University}

\abstract{Many rare weather events, including hurricanes, droughts, and floods, dramatically impact human life. To accurately forecast these events and characterize their climatology requires specialized mathematical techniques to fully leverage the limited data that are available. Here we describe \emph{transition path theory} (TPT), a framework originally developed for molecular simulation, and argue that it is a useful paradigm for developing mechanistic understanding of rare climate events. TPT provides a method to calculate statistical properties of the paths into the event. As an initial demonstration of the utility of TPT, we analyze a low-order model of sudden stratospheric warming (SSW), a dramatic disturbance to the polar vortex which can induce extreme cold spells at the surface in the midlatitudes. SSW events pose a major challenge for seasonal weather prediction because of their rapid, complex onset and development. Climate models struggle to capture the long-term statistics of SSW, owing to their diversity and intermittent nature. We use a stochastically forced Holton-Mass-type model with two stable states, corresponding to radiative equilibrium and a vacillating SSW-like regime. In this stochastic bistable setting, from certain probabilistic forecasts TPT facilitates estimation of dominant transition pathways and return times of transitions. These ``dynamical statistics" are obtained by solving partial differential equations in the model's phase space. With future application to more complex models, TPT and its constituent quantities promise to improve the predictability of extreme weather events, through both generation and principled evaluation of forecasts.}

\begin{document}

\maketitle

%
%
%

%








\section{ Introduction and background}
The polar winter stratosphere typically supports a strong, cyclonic polar vortex, maintained by the thermal wind relation and meridional temperature gradient. A sudden stratospheric warming (SSW) event is a large excursion from this normal state, which can take many different forms. In split-type SSWs, the vortex splits completely in two. In displacement-type SSWs the vortex displaces far away from the pole (these can be considered wavenumber-2 and wavenumber-1 disturbances, respectively). Both types are ``major'' warmings, as the mean zonal wind reverses. In a ``minor'' warming, the zonal wind slows down significantly without completely reversing \citep{butler}. 

SSW is a rare event occurring about twice every three years, depending on the definition used \citep{butler}. Its effects can propagate downward into the troposphere, altering the tropospheric jetstream and inducing extreme midlatitude surface weather events, including cold spells and precipitation \citep{baldwin_dunkerton,thompson_baldwin}. Abrupt cold spells severely stress infrastructures, economies and human lives, and every bit of extra prediction lead time is helpful for adaptation. Unfortunately, numerical weather prediction struggles to forecast SSW at any lead time longer than about two weeks \citep{multiple_nwp}. Understanding SSW is therefore important for practical forecasting as well as science, but this task remains difficult. Several different geophysical fields are often used as indices of SSW onset. One simple indicator is zonal-mean zonal wind at $60^\circ$N, which defines thresholds for minor and major warming \citep{cp07,butler}. Another common indicator is the 10hPa geopotential height field, which was used by \cite{nfdr} to estimate a fluctuation-dissipation relation in its leading empirical orthogonal functions (EOFs). Many studies have examined SSW precursors and dominant pathways through simulation and observation. \cite{lifecycle}, for instance, catalogued the various wavenumber forcings, heat fluxes and zonal wind anomalies that accompanied each stage of SSW events from reanalysis data. While planetary wave forcing from the troposphere is an accepted proximal cause of SSW, the polar vortex's susceptibility to such forcing, or ``preconditioning'', is a nontrivial and debated function of its geometry \citep{albers_birner, bancala}. Tropospheric blocking is also thought to be linked to SSW; \cite{martius} and \cite{bao} found blocking to precede many major SSW events of the past half century. The diversity and complex life cycle of SSWs makes it difficult to build a unified picture of their onset.

\begin{figure}[t]
    \caption{The committor function for a double-well potential under the dynamics $\dot x=-V'(x)+\sigma\dot w$. Panel (a) shows the potential function $V(x)$, and panel (b) shows the committor function. The committor has value zero on the left minimum, one on the right minimum, and one half at the top of the barrier. The stronger the stochastic forcing, the less the actual potential shape matters and the more gradual the committor's slope. For small noise, the dynamics become more deterministic and the committor approaches a step function, since $x(t)$ will directly approach whichever minimum is closer.}
    \includegraphics[width=\linewidth]{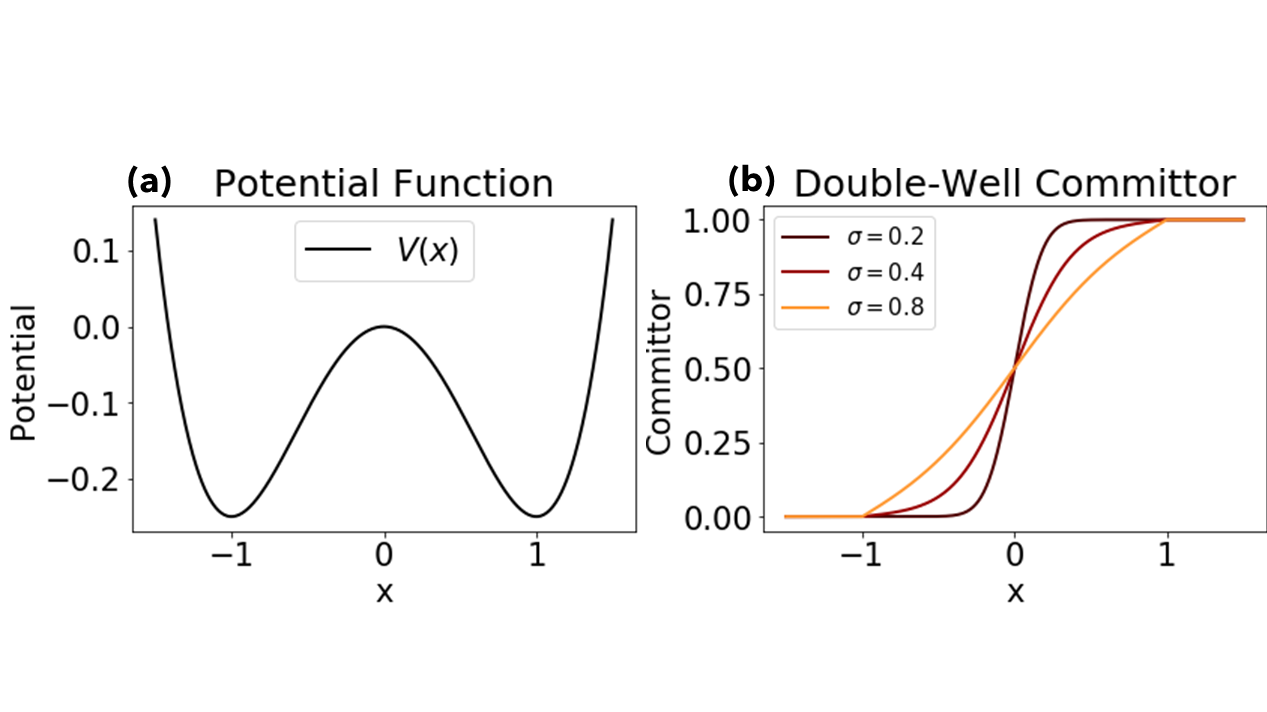}
    \label{fig:doublewell}
\end{figure}

In this article, we are interested in developing a detailed understanding of transition events between two states, at least one of which is typically long-lived. Consider, for example, a particle with position $x(t)$ moving in the double well potential energy landscape $V(x)=\frac{x^4}{4}-\frac{x^2}{2}$ (illustrated in Figure \ref{fig:doublewell}) and forced by stochastic white noise $\dot W$: $\dot x=-V'(x)+\sigma\dot W$. If the system starts in the left well, it will tend to remain there a while, but occasionally the stochastic forcing will push it over the barrier into the right well. The natural predictor for this event is the \emph{committor}: the probability of reaching the right well before the left well.

We denote this function by $q(x)$, which solves the Kolmogorov backward equation (to be introduced later). For this simple system the equation takes a form which can be solved exactly:
\alnn{
    \begin{cases}
    -V'(x)q'(x)+\frac{\sigma^2}2q''(x)=0 & x\in(-1,1)\\ 
    q(-1)=0,\ \ q(1)=1 \\
    \end{cases}\\
    \implies q(x)=\frac{\int_{-1}^x\exp\big(\frac2{\sigma^2}V(x')\big)\,dx'}{\int_{-1}^1\exp\big(\frac2{\sigma^2}V(x')\big)\,dx'}
}
Note that the boundary conditions are implied by the probabilistic interpratation.

The committor is plotted in the right panel of Figure \ref{fig:doublewell} for various noise levels. N.B., the potential landscape picture is not fully general, but is a useful mental model. The equations that determine $q(x)$ will be presented in section 2.

In the case of SSW, the long-lived states are the steady and disturbed circulation regimes of the stratospheric polar vortex. Recent work by \cite{bouchet} has studied SSW in an equilibrium statistical mechanics framework, with these two stable states as saddle points of energy functionals. TPT takes a complementary non-equilibrium view, describing the long-time (steady-state) statistics of trajectories between the two states. For example, TPT introduces a probability density of reactive trajectories (or ``reactive density") indicating the regions where trajectories tend to spend their time en route from A to B. The system is said to be \emph{reactive} at a point in time if it has most recently visited $A$ and will next visit $B$. The associated probability current of reactive trajectories (or ``reactive current") indicates the preferred direction and speed of transition paths. These detailed descriptors of the mechanism underlying a rare event can be expressed in terms of probabilistic forecasts like the committor $q(x)$, the probability of entering state A before reaching state B from a given initial condition $x$ (not in either A or B). The committor is the ideal probabilistic forecast in the usual variance-minimizing sense of conditional expectations \citep{Durrett}. Any other predictor of a transition derived through experiments and observations, such as vortex preconditioning and forcing at different wavenumbers \citep{albers_birner,bancala,martius,bao} necessarily corresponds to an approximation of the committor. 

Ensemble simulation is a commonly used method to estimate the committor at a single given initial condition by measuring the fraction of ensemble trajectories that achieve the rare event. This is challenging because many simulations are needed to generate enough rare events for significant statistical power. Recent and ongoing work aims to channel this computing power more efficiently in weather simulation using importance sampling and large deviation theory \citep{Hoffman2006,Weare2009,VandenEijnden2013,Ragone24,rogue,plotkin,webber}. The committor, and other quantities of interest, are fundamentally averages over sample trajectories. We can also express them as solutions to a concrete set of partial differential equations (PDEs) using basic stochastic calculus. TPT provides a framework to exploit these quantities and enhance our understanding of rare events, both from simulation data and from the fundamental equations of motion.

TPT has been applied primarily to molecular dynamics simulation to determine reaction rates and pathways of complex conformational transitions \citep{pathfinding,tpt_simple_examples,towards}; however, the framework does not depend on the details of the underlying dynamical system. TPT deals particularly with stochastically forced systems, such as Brownian dynamics of particles. Stochastic forcing applies quite generally; while the climate system is deterministic in principle, nonlinear interactions between resolved and unresolved scales inevitably leads to resolution-dependent model errors that can be approximated as stochastic. \cite{hasselman} originally formulated stochastic climate models to capture the influence of quickly evolving ``weather" variables on the slowly evolving ``climate" variables. Stochastic parameterization remains an active area of research. For example, \cite{franzke_majda2006} had success in capturing energy fluxes of a 3-layer quasigeostrophic model by projecting onto ten EOF modes and treating the remainder as stochastic forcing. \cite{kitsios_frederiksen} addressed the challenge of designing consistent numerical schemes for subgrid-scale parametrization. Deep convection in the atmosphere and turbulence in the ocean boundary layer are two examples of multiscale processes that are especially challenging to resolve. 

The aim of this article is to introduce the key quantities and relations describing the path properties of rare atmospheric events.  Computing those quantities for more complicated systems than the low order model studied here is a significant and, we argue, worthwhile challenge.  We do not address that computational challenge here.  Instead we note that development of approximation techniques for TPT and related quantities in high dimensional settings is an active area of research \cite{dga,Bowman2009,Chodera2014}.  These methods incorporate both simulated and observational data and we outline them briefly in the conclusion. 

The paper is organized as follows. Section 2 describes the dynamical model we use, building on work by \cite{ruz} and \cite{birner_williams}. Section 3 describes the mathematical framework of TPT, with detailed, but informal, derivations mainly put in the supplement. Section 4 explains the methodology and results particular to this model. 

\section{ Dynamical model}
\cite{holton_mass} studied ``stratospheric vacillation cycles,'' a certain kind of minor warming in which zonal wind oscillates on a roughly seasonal timescale. They posited a mechanism of  wave-mean flow interaction, which continues to be an important modeling paradigm. The quasi-geostrophic equations are confined to a $\beta$-plane channel from $60^\circ$ to the north pole, and the streamfunction is perturbed from below by orographically induced planetary waves, specified through the lower boundary condition. The Holton-Mass model combines the zonal-mean flow equations,

\begin{align} 
\pder{\ov u}{t}-f_0\ov v&=0 \label{eq:qg1} \\
f_0\ov u&=-\pder{\ov\Phi}{y} \label{eq:gq2} \\
\pder{}{t}\bigg(\pder{\ov\Phi}{z}\bigg)+N^2\ov w&=-\alpha\bigg(\pder{\ov\Phi}{z}-\frac{R\ov T\st}{H}\bigg)-\pder{}{y}(\ov{v'\Phi_z'}) \label{eq:qg3} \\
\pder{\ov v}{y}+\frac1{\rho_s}\pder{}{z}(\rho_s\ov w)&=0 \label{eq:qg4}
\end{align}
and the linearized quasigeostrophic potential vorticity equation,
\begin{align}
    \bigg(\pder{}{t}+\ov u\pder{}{x}\bigg)q'&+\beta_e\pder{\psi'}{x}+\frac{f_0^2}{\rho_s}\pder{}{z}\bigg(\frac{\alpha\rho_s}{N^2}\pder{\psi'}{z}\bigg)=0\\
    \text{where }\ \ \ \  q'&:=\nabla^2\psi'+\frac1{\rho_s}\pder{}{z}\bigg(\frac{f_0^2}{N^2}\rho_s\pder{\psi'}{z}\bigg)\\
    \text{and }\ \ \ 
    \beta_e&=\beta-\pdersq{\ov u}{y}-\frac1{\rho_s}\pder{}{z}\bigg(\rho_s\frac{f_0^2}{N^2}\pder{\ov u}{z}\bigg).
\end{align}

These equations represent (2) zonal and (3) meridional momentum balance, (4) conservation of energy, (5) conservation of mass, and (6) a combination of all these for the perturbations. Overbars and primes represent zonal averages and perturbations. $\Phi$ is the geopotential height; $H=7\,km$ is the atmospheric scale height; $z=-H\ln(\frac{p}{p_0})$ is log-pressure; $\rho_s=\rho_0e^{-z/H}$ is a standard density profile; $\alpha=\alpha(z)$ is an altitude-dependent damping coefficient; and $\ov T\st=\ov T\st(y,z)$ is the radiative equilibrium temperature. \cite{holton_mass} studied the solution consisting of single Fourier modes in the zonal and meridional directions:
\begin{align}
\ov u(y,z,t)&=U(z,t)\sin\ell y \label{ansatz1}\\
\psi'(x,y,t)&=\re\{\Psi(z,t)e^{ikx}\}e^{z/2H}\sin\ell y \label{ansatz2}
\end{align}
where $\psi'$ is the zonal perturbation of $\psi=(g/f_0)\Phi$. $k=2/(a\cos60^\circ)$ and $\ell=3/a$ are the zonal and meridional wavenenumbers, where $a$ is Earth's radius. These wavenumbers are commonly observed in real SSWs and used in theoretical studies \citep{birner_williams,ruz,yoden,holton_mass}; a split-type SSW is an extreme wavenumber-two perturbation.  
The lower boundary condition at $z=z_B$ (the tropopause) is
\begin{align}
    \Psi(z_B,t)=\frac{g}{f_0}h(t)
\end{align}
where $h$ is a topographically induced perturbation to geopotential height at the tropopause. Holton and Mass found that for a certain range of $h$, this system has qualitatively different regimes: a steady eastward zonal flow close to radiative equilibrium, and a weaker zonal flow with quasi-periodic ``vacillations'' from eastward to westward, even under constant forcing. Each vacillation cycle consists of a sudden warming and cooling over the timescale of weeks. Although these individual cycles are interesting weather events unto themselves, in this paper we think of the vacillations as occurring within a general \emph{climate regime} that is conducive to sudden warming, as opposed to the steady flow state, which is not. Transitions between these two regimes, which we focus on here, are more accurately described as climatological shifts than weather events. The study by \cite{ruz} varies $h$ on an interannual timescale, with each single winter season occupying one of the two stable states and generating its daily weather accordingly. Hence, for this paper we will use the term ``climate transitions." 

The original Holton-Mass model discretizes the above PDE with finite differences across 27 vertical levels, which is assumed to be close to a continuum limit. Following several studies at this resolution \citep{holton_mass,yoden,Christiansen2000}, \cite{ruz} did the most severe truncation possible, resolving only three vertical levels (including fixed boundaries) for easy analysis and exploration of parameter space. This reduces phase space to only three degrees of freedom: 
$U(t)$, which modulates $\ov u$ as a sine jet; $X(t)=\re\{\Psi(t)\}$; and $Y(t)=\im\{\Psi(t)\}$. $X$ and $Y$ modulate the amplitude and phase of the perturbation streamfunction:
\begin{align}
\psi'(x,y,t)&=(X\cos kx-Y\sin kx)e^{z/2H}\sin\ell y 
\end{align}

Carrying the Ansatz through the QG equations, \cite{ruz} derived the following system:
\begin{align}
\dot X&=-\frac1{\tau_1}X-rY+sUY-\xi h+\delta_w\dot h \label{eqn:X}\\
\dot Y&=-\frac1{\tau_1}Y+rX-sUX+\zeta hU \label{eqn:Y}\\
\dot U&=-\frac1{\tau_2}(U-U_R)-\eta hY-\delta_\Lambda\dot\Lambda \label{eqn:U}
\end{align}

The primary control parameter, $h$, represents topographic forcing and other sources of planetary waves, such as land-sea ice contrast. While \cite{ruz} also varies $\Lambda$, representing vertical wind shear, we will only vary $h$ and set $\Lambda$ constant. Time derivatives $\dot h$ and $\dot\Lambda$ are zeroed, removing transient forcings such as seasonality effects. Appendix A and \cite{ruz} have a detailed list of parameters.

\begin{figure}[t]
    \caption{Fixed points of Equations (\ref{eqn:X})-(\ref{eqn:U}) in the state space $(X,Y,U)$, where $X$ and $Y$ represent the real and imaginary parts of the streamfunction and $U$ the mean zonal wind amplitude. Fixed points vary as a function of the topographic forcing parameter, $h$. Panels (a), (b) and (c) show fixed points of $X$, $Y$ and $U$ respectively on the vertical axis, while $h$ varies across the horizontal axis. Circles and crosses denote linearly stable and unstable fixed points, respectively. The range of $h$ between $\sim20\,m$ and $\sim160\,m$ supports three fixed points, two stable and one unstable. In this range, the blue points correspond to the radiative solution, while the red points represent the vacillating regime. This corresponds to a winter climatology that is conducive to sudden stratospheric warming events.}
    \centering
    \includegraphics[width=\linewidth]{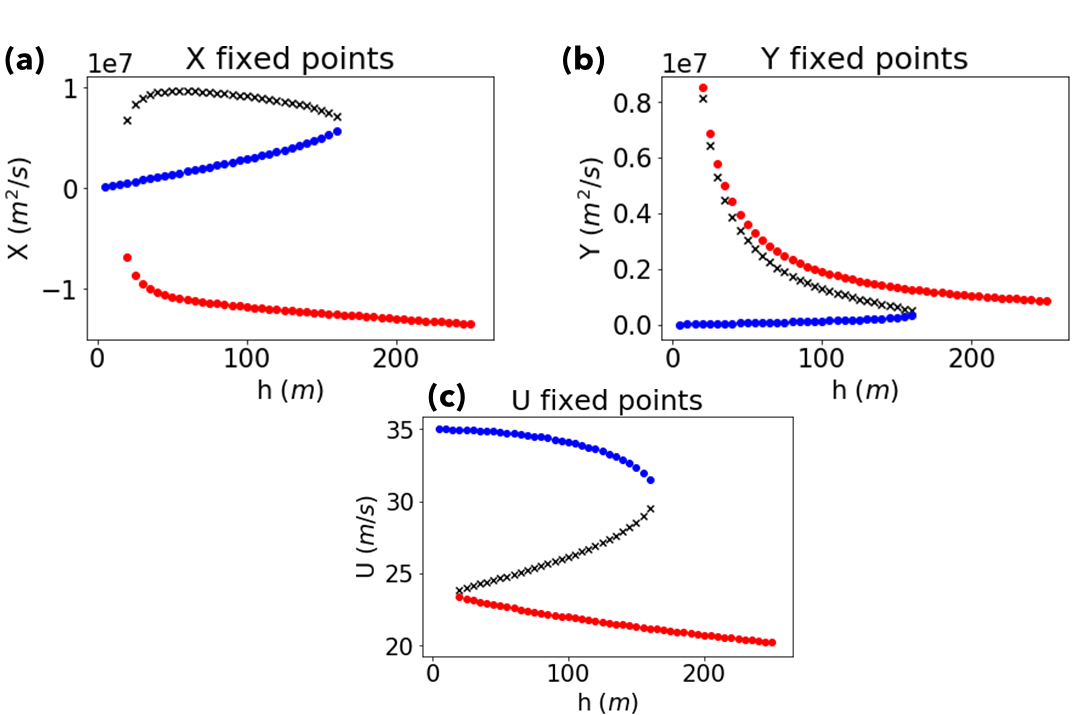}
    \label{fig:bif}
\end{figure}

\begin{figure}
    \caption{Trajectories in $(X,Y,U)$ space, where $X$ and $Y$ represent the real and imaginary parts of the streamfunction and $U$ the mean zonal wind amplitude. In this simulation, the topographic forcing $h$ increases linearly from $0\,m$ to $200\,m$ in 1300 days. (a) shows the fixed points, with colors blue, red and black for the radiative solution ($A$), the vacillating solution ($B$) and the unstable fixed point between them respectively. The trajectory of $U$ over time is superimposed in gray. (b) plots this same curve parametrically, in $XU$ space. Before the bifurcation, the trajectories follow the existing fixed point; after the bifurcation, they spiral into the new fixed point through a series of ``vacillations."}
    \includegraphics[width=\linewidth]{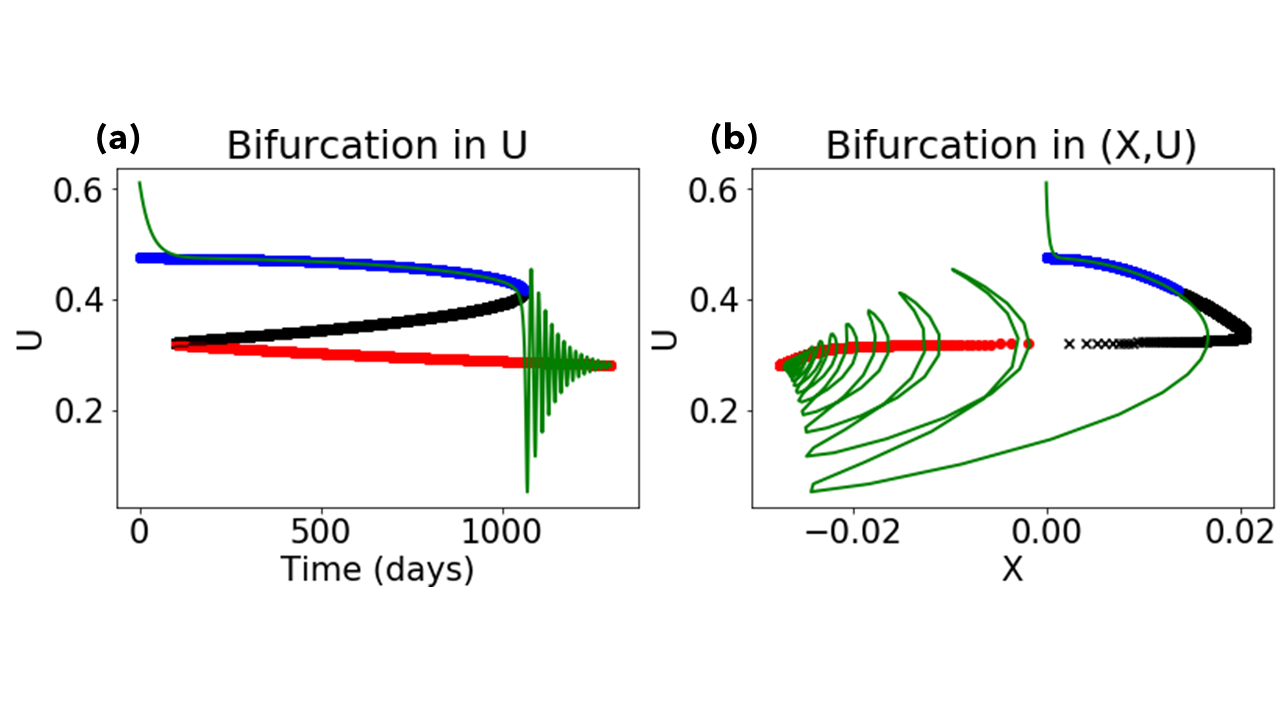}
    \label{fig:cross_bif}
\end{figure}

Remarkably, this hugely simplified model retains the qualitative structure of the Holton-Mass model as a bistable system for a certain range of $h$ between the critical values $h_1\approx20\,m$ and $h_2\approx160\,m$, as shown in the bifurcation diagram of Figure \ref{fig:bif}. Blue points represent the normal state of the vortex, in approximate thermal wind balance with the radiative equilibrium temperature field (henceforth called the ``radiative solution"). Red points represent a disturbed vortex, with weaker zonal wind and vacillations. This climatological regime supports more SSW events, and is henceforth called the ``vacillating solution." We use the same blue-red color scheme consistently here to represent these two states. Transitions between them happen on interannual time scales, affecting each year's likelihood of SSW events. The structure of transitions is illustrated in Figure \ref{fig:cross_bif}: as $h$ increases slowly past the bifurcation threshold $h_2$, the system enters a series of rapid, large-amplitude oscillations that spiral into the weaker-circulation state.  

\begin{figure}
    \caption{Stochastic trajectories of the system for various fixed values of the parameters $h$ (topographic forcing) and $\sigma_3$ (amplitude of stochastic forcing). Panels (a), (b) and (c) show $U(t)$ for three different forcing levels: $h=25,35,45\,m$ with $\sigma_3=0.5\,m/s/day^{1/2}$. In keeping with the bifurcation diagrams, the blue, black and red lines mark the radiative, unstable and vacillating solutions respectively. Note that their relative positions vary slightly with $h$, as fixed points depend on parameters. As $h$ increases from left to right, the systems spends increasingly more of its time in the vacillating state. Panel (d) shows a parametric plot of the transitions through $(X,U)$ space, for $h=35\,m$ (another view of panel (b)). The transition happens seven times, and hence panel (d) shows seven different transition paths superimposed on each other. Most of the transitions follow a similar characteristic curve through $XU$ space, with a rapid decrease in $U$ followed by a decrease in $X$.}
    \centering
    \includegraphics[width=\linewidth]{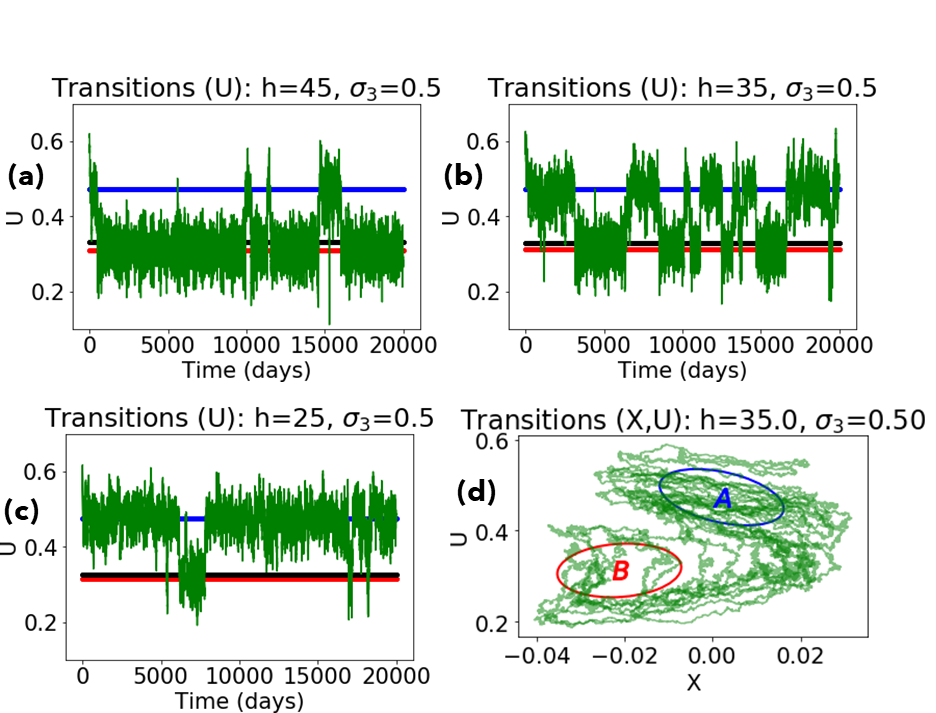}
    \label{fig:path_samp}
\end{figure}

In Figures \ref{fig:bif} and \ref{fig:cross_bif}, transitions require crossing the bifurcation threshold $h_2$, where the radiative solution ceases to exist. \cite{birner_williams} introduced additive white-noise forcing in the $U$ variable to model unresolved gravity waves and found that these perturbations were sufficient to excite the system out of its normal state and into a vacillating regime. In Figure \ref{fig:path_samp} we illustrate stochastic trajectories of the system for three different (fixed) values of $h$. (For numerical reasons we also add a small amount of independent white noise to $X$ and $Y$ variables). Even when $h$ is far below $h_2$, transitions still occur, and in fact the preference for the vacillating solution branch increases quickly with $h$.

\cite{birner_williams} used direct numerical simulation and the Fokker-Planck equation to calculate long-term occupation statistics, i.e., how much time on average was spent in each regime and the mean first passage time before a transition to the vacillating regime, all for a range of forcing and noise levels. Our approach differs in both target and methodology. We aim to characterize the transition process between the two states, to monitor its progress in real time, as well as to describe statistics of the transition over many realizations. Methodologically, transition path theory phrases these questions in terms of the \emph{generator} of the stochastic process, a differential operator that encodes all information about the behavior of the process. 

\section{ Path properties}
TPT characterizes the steady state statistics of transitions between states. In this section, we introduce key quantities needed to introduce TPT as applied to the Ruzmaikin model to obtain a more complete picture than we get from the sample paths shown above. For mathematical details, see the supplement and background literature \citep{towards,tpt_simple_examples,pathfinding}.

\subsection*{a. Infinitesimal Generator}

The noisy Ruzmaikin model can be expressed compactly as a stochastic differential equation (SDE)---specifically a \emph{diffusion} process---in the variable $Z=(X,Y,U)\in\R^3$ (or a more general phase space $\Omega$) with a deterministic drift vector $b(z)=\big(b_1(z),b_2(z),b_3(z)\big)$ and a $3\times3$ diffusion matrix $\sigma(z)$.
\alnn{
dZ_t=b(Z_t)\,dt+\sigma(Z_t)\,dW_t
}
Here, $W_t$ is a 3-vector of independent Brownian motions. We use the Ito convention for stochastic integration. While $\sigma$ can in principle be any $z$-dependent matrix, we make $\sigma$ diagonal and constant: $\sigma(z)=\diag(\sigma_1,\sigma_2,\sigma_3)$, creating independent additive noise in the $X$, $Y$ and $U$ variables. $\sigma_1$ and $\sigma_2$ have units of $m^2/s/day^{1/2}$, while $\sigma_3$ has units of $m/s/day^{1/2}$. Associated with this equation is the infinitesimal generator $\lcal$, an operator describing the evolution of observable functions forward in time following a trajectory. Explicitly, if $f(\cdot)$ is a smooth function of phase space variables, then
\alnn{
\lcal f(z)&:=\frac{d}{dt}\ex[f(Z_t)|Z_0=z]\Big|_{t=0}
}
where $\ex$ is an expectation over sample paths. 

Ito's lemma (the chain rule for diffusion SDEs) gives the Kolmogorov backward equation, which represents $\lcal$ as a partial differential operator (for details see \cite{pavliotis}):
\alnn{
\lcal f(z)&=\sum_ib_i(z)\pder{f(z)}{z_i}+\frac12\sum_{i,j}(\sigma\sigma\tr)_{ij}\frac{\partial^2f(z)}{\partial z_i\partial z_j}\\
&=b(z)\cdot\nabla f(z)+\text{Tr}\Big(\frac12\sigma\sigma\tr Hf(z)\Big)
}

The diffusion matrix $\frac12\sigma\sigma\tr$ is also called $D$ for convenience, which we will use interchangeably. $Hf$ denotes the Hessian matrix: $[Hf]_{ij}=\partial^2f/\partial z_i\partial z_j$.  The generator provides path statistics as the solution to PDEs, as illustrated in the following subsections.

\subsection*{b. Equilibrium probability density}
This stochastic process admits a time-dependent probability density, $\rho(z,t)$, which can be derived from the generator. For example, if the system starts in a known position $Z_0=z$, then $\rho(z',0)=\delta(z-z')$. The density spreads out from this initial point over time according to the Fokker-Planck equation, which can be written in terms of the adjoint of the generator:
\alnn{
&\pder{\rho(z,t)}{t}=\lcal\st\rho(z,t)\\
&=\sum_i\pder{}{z_i}\Big[-b_i(z)\rho(z,t)+\sum_j\pder{}{z_j}\big(\rho(z,t)D_{ij}(z)\big)\Big]\\
&=\nabla\cdot\big[-b(z)\rho(z,t)+\nabla\cdot\big(\rho(z,t)D(z)\big)\big]
}
When $D$ is constant and diagonal, the last term simplifies to $\nabla\cdot\big[\nabla\cdot(\rho D)\big]=\sum_iD_{ii}\partial^2\rho/\partial z_i^2$. In the case of pure Brownian motion, $dZ_t=dW_t$, then $b=0$ and $D=\frac12I$, giving the heat equation $\partial_t\rho=\frac12\nabla^2\rho$ ($\nabla^2=\sum_i\frac{\partial^2}{\partial z_i^2}$). Assuming that the process is ergodic, the density eventually forgets the initial condition and stabilizes into a long-term (or equilibrium, or stationary) probability density $\pi(z)$, which solves $\lcal\st\pi=0$. This can be approximated by either simulating the SDE for a very long time and binning data points, or directly solving the stationary PDE $\lcal\st\pi=0$, subject to the normalization and positivity constraints $\int_\Omega\pi(z)\,dz=1$ and $\pi(z)\geq0$. 

\subsection*{c. Committor probability}
The stationary density is an equilibrium quantity characterizing the long-term occupation statistics. But it is insufficient to describe the events of interest to us, which are \emph{transition paths}: trajectory segments beginning inside the radiative state and ending inside the vacillating state. Specifically, we define the sets $A$ and $B$ as ellipsoids around these two fixed points, respectively. Their size is determined by contours of a local approximation to the stationary density $\pi$; see supplement for details. We say that a snapshot $Z_t$ of the system is undergoing a \emph{transition} (or reaction) at time $t$ if it is on the way from set $A$ to set $B$. This involves information about both its future and its past, for which we introduce the forward and backward committor probabilities in this section. 

The forward committor $\qp$ (denoted $q$ when context is clear) describes the progress of a stochastic trajectory traveling from set $A$ to set $B$, as follows:
\alnn{
\qp(z)&=\pr\{Z_t\text{ next hits }B\text{ before }A|Z_0=z\} \\
 &\qp(z\in A)=0, \ \ \ \ \ \qp(z\in B)=1 
}
The boundary conditions on $A$ and $B$ follow naturally from the probabilistic definition. If the system begins in set $A$, by path continuity it will certainly next find itself in $A$, with zero chance of hitting $B$ first. Starting in set $B$ the opposite is true. The committor therefore obeys the boundary value problem (see supplement for derivation)
\alnn{
\begin{cases}
\lcal\qp(z)=0 & z\in(A\cup B)^c\\
\qp(z)=0 & z\in A\\
\qp(z)=1 & z\in B\\
\end{cases}
}
The equivalence of conditional expectations with respect to a Markov process like the committor and solutions to PDE involving the generator of the process are generally referred to as Feynman-Kac relations \citep{Karatzas} and are well studied. The PDE in (25) is most naturally posed on an infinite domain, but as a numerical approximation we solve it in a large rectangular domain and impose homogeneous Neumann conditions at the domain boundary. 
A limiting example is the noise-dominated case, where $b(z)$ is negligible and $D=I$. The Kolmogorov backward equation then becomes Laplace's equation: 
\alnn{
\lcal\qp(z)=\nabla^2\qp(z)=0
}
If posed on the interval $[0,1]$, with $A=\{0\}$ and $B=\{1\}$, the solution is $q^+(z)=z$. The linear increase from set $A$ to set $B$ reflects the greater likelihood of entering $B$ when beginning closer to it. This limit is reflected in Figure \ref{fig:doublewell}, which shows the committor of the double-well potential approaching a straight line for large $\sigma$ values.  

Prediction is naturally much harder in high-dimensional systems such as stratospheric models. A number of physically interpretable fields, such as zonal wind and geopotential height anomalies, seem to have some predictive power for SSW, but prediction by any single such diagnostic is suboptimal. Insofar as they are successful, these variables approximate certain aspects of the committor. For example, the committor might increase monotonically with the quasi-biennial oscillation index. Furthermore, statistical correlations potentially obscure the conditional relationships needed. For example, \cite{martius} and \cite{bao} examined tropospheric precursors to SSW events in reanalysis records, finding that blocking events preceded most major SSWs, potentially by enhancing upward-propagating planetary waves. (We use ``precursor" only to mean an event that sometimes happens before SSW.) Blocking influences SSW through height perturbations at the tropopause, which would enter the Ruzmaikin model as low-frequency variations in lower boundary forcing $h$. Since we fix $h$ constant, the blocking precursor is outside our scope here. However, farther down the dynamic chain are other measurable precursors such vertical wave activity flux and meridional heat flux, which are also found to have predictive power \citep{Sjoberg2012}. 
However comprehensive the model, we would naturally expect the true committor probability to exhibit similar patterns to canonical precursors of that model such as blocking (for a troposphere-coupled model) and heat flux (for a stratosphere-only model). Yet, there is an important difference: while a precursor $P$ may appear with high probability \emph{given} that a SSW is imminent, the committor specifies the probability of a SSW given an observed pattern. As acknowledged in \cite{martius}, many blocking events did not lead to SSW events, meaning that $\pr\{\text{SSW}|\text{blocking}\}\neq\pr\{\text{blocking}|\text{SSW}\}$. Such distinctions highlight the need for a precise mathematical formulation that provides and distinguishes both kinds of information.

While $\qp$ describes the future of a transition, the backward committor $\qm$ describes its past. It is defined as
\alnn{
\qm(z)&:=\pr\{Z_t\text{ last visited }A\text{ rather than }B|Z_0=z\}\\
&\qm(A)=1,\ \ \ \ \ \qm(B)=0
}
$\qm$ solves the time-reversed Kolmogorov backward equation $\wt\lcal\qm=0$, where $\wt\lcal$ is the time-reversed generator, which evolves observables backward in time. See supplement for a detailed description of $\wt\lcal$ and its relationship the forward generator $\lcal$. 

We now describe the fundamental statistics characterizing transition events as identified by TPT and explain how they can be expressed in terms of quantities such as $\qp$, $\qm$, and $\pi$. The probability density of reactive trajectories $\rho_R(z)$, the probability of observing the system $Z_t$ at the location $z$ during a transition, is proportional (up to a normalization constant) to the product $\pi(z)\qm(z)\qp(z)$. This density is large in regions of phase space that are highly trafficked by reactive trajectories. This is how TPT gives information about precursors, indicating regions of phase space that are usually visited by the system over the course of a transition path.

The direction and intensity of this traffic is specified by the \emph{reactive current}. To develop this concept, we start by introducing the probability current $J$, a vector field that satisfies a continuity equation with the time-dependent density $\rho$: 
\aln{
\pder{\rho}{t}=\lcal\st\rho=-\nabla\cdot J
}
If $\rho$ were the density and $v$ the velocity field of a fluid, $J$ would be $\rho v$. One can think of $J$ as an instantaneous (in time and position) average over all possible system trajectories, though a precise mathematical description requires some care. In equilibrium, when $\rho=\pi$ is no longer changing, $\nabla\cdot J=0$, or equivalently $\oint_CJ\cdot n\,dS=0$ where $C$ is any closed surface. 

The reactive current $J_{AB}$ is also an ``average velocity", but restricted to reactive paths. Unlike $J$, $J_{AB}$ is not divergence-free, with a source in $A$ and a sink in $B$ (where transition paths start and end). $J_{AB}$ is defined  implicitly via surface integrals. If $C$ is any surface enclosing set $A$ but not set $B$, with outward normal $n$, then the flux $\oint_CJ_{AB}\cdot n\,dS$ is the number of forward transitions per unit time, also called the \emph{transition rate} $R_{AB}$; see \citet{pathfinding} for details. The supplement describes another expression in terms of the generator. The result is \citep{tpt_simple_examples}
\alnn{
J&=\pi b-\nabla\cdot(\pi D) \\
J_{AB}&=\qp\qm J+\pi D(\qm\nabla\qp-\qp\nabla\qm) 
}
where again $\pi$ is the stationary density. This expression has intuitive ingredients. Multiplying $J$ by $\qp\qm$ conditions the equilibrium probability current on the trajectory being reactive, meaning en route from $A$ to $B$. The $\qm\nabla\qp-\qp\nabla\qm$ reflects the fact that trajectories from $A$ to $B$ must ascend a gradient of $\qp$, going from $\qp=0$ to $\qp=1$, while descending a gradient of $\qm$.

Just as $J_{AB}(z)$ describes the average reactive velocity, a streamline $z_t$ of $J_{AB}(z)$ (solving $\dot z_t=J_{AB}(z)$, with $z_0=a\in A$ and $z_T=b\in B$ for some $T>0$) is a kind of ``average" transition path. Although the streamline will not be realized by any particular transition path, it will have common geometric features in phase space with many actual path samples. At low noise the reactive trajectories will cluster in a thin corridor about the streamline. The streamline is a more dynamical description of precursors: whereas regions of high reactive density are commonly observed states along reactive trajectories, streamlines of reactive current are commonly observed \emph{sequences} of states along reactive trajectories. The study by \cite{lifecycle}, for example, described a sequence of events in a prototypical SSW life cycle based on reanalysis including vortex preconditioning, wave forcing, and anomalous heat fluxes at various levels in the troposphere and stratosphere. The sequence described there likely corresponds to a streamline of the reactive trajectory.

The committor also quantifies the relative balance of time spent on the way to each set. If more probability mass lies in the region where $\qp>\frac12$, set $B$ is globally more imminent, whereas more mass where $\qp<\frac12$ indicates set $A$ is. A single summary statistic of imminence is the average committor during a long trajectory, $\ex[\qp(Z_t)]$, computed as a weighted average against the equilibrium density:
\alnn{
\ex[\qp(Z_t)]=\ex_\pi[\qp(Z)]=\int_\Omega\qp(z)\pi(z)\,dz=:\langle\qp\rangle_\pi
}
An average below (above) $1/2$ would indicate more time spent on the way to to $A$ ($B$). 

Another statistic, the forward transition rate, captures the frequency of transitions between $A$ and $B$ rather than the overall time spent in each. We earlier defined $R_{AB}$ as the number of $A\to B$ transitions per unit time. Since a $B\to A$ transition must occur between every two $A\to B$ transitions, $R_{AB}=R_{BA}=:R$. The inverse of the transition rate is the return time, a widely used metric for changing frequency of extreme events under climate change scenarios \citep{easterling}. However, the forward and backward transitions may differ in important characteristics like speed. To capture this asymmetry, we need a \emph{dynamical} analogue to the equilibrium statistic $\langle\qp\rangle_\pi$. The typical quantity of choice is the rate \emph{constant} $k_{AB}$, which is larger if $A\to B$ transitions happen faster than $B\to A$ transitions. We therefore normalize by the overall time spent having come from $A$, which is $\langle\qm\rangle_\pi$. 
\alnn{
k_{AB}=\frac{R}{\langle\qm\rangle_\pi}=\frac{R}{\int\qm(z)\pi(z)\,dz}
} 
This rate constant, defined in \cite{rate}, parallels the chemistry definition. If $X_A$ and $X_B$ are two chemical species, with $[\cdot]$ denoting concentration, the forward and backward rate constants $k_{AB}$ and $k_{BA}$ are defined so that
\alnn{
R=[X_A]k_{AB}=[X_B]k_{BA}
}
In the language of transition path theory, $[X_A]$ is the long-term probability of the system existing most recently in state $A$, which is $\langle\qm\rangle_\pi$. Rates are also expressible in terms of expected passage times. Thinking of $[X_A]$ as the total probability of having last visited set $A$, $1/k_{AB}=[X_A]/R$ estimates the total transition time between entering $A$ (having last visited $B$) and next re-entering $B$. It is these inverse quantities we display in the results section. 

These quantities together make an informative description of the typical transition process from $A$ to $B$. We now proceed to analyze the transition path properties of the Ruzmaikin stratospheric model.

\section{Methodology}

\subsection*{a. Spatial discretization}
The quantities of interest described above ($\pi$, $\qp$, $\qm$, and $J_{AB}$) emerge as solutions to PDEs involving the generator $\lcal$, which must be approximated by spatial discretization. In the supplement we describe a finite volume scheme to directly discretize the adjoint $\lcal\st$ as a matrix, which we name $L\st$, on a regular grid in $d$ dimensions. Here we use the same domain and noise levels as \cite{birner_williams}: $-0.06\leq X\leq0.04$, $-0.05\leq Y\leq0.05$, $0\leq U\leq0.8$ in units non-dimensionalized in terms of the radius of Earth and the length of a day. We tile this with a grid of $40\times40\times80$ grid cells. We choose a noise constant $\sigma_3$ in the $U$ variable in the range $0.4-1.5$. This is a similar range to observed atmospheric gravity wave momentum forcing \citep{birner_williams}. For numerical reasons, we also add small noise to the streamfunction variables $X$ and $Y$, in proportion to the domain size. Specifically, as $U$ spans a range of $0.8$ and $X,Y$ span a smaller range of $0.1$, we choose $\sigma_1$ and $\sigma_2$ to be $\sigma_3\times(0.1/0.8)$. This adjustment does change our results with respect to \cite{birner_williams}, causing more transitions in both directions at lower $h$ than if only the $U$ variable were perturbed. While gravity wave drag forces the zonal wind, eddy interactions and other sources of internal variability can perturb the streamfunction as well, and it is not uncommon to represent these effects stochastically \citep{Delsole1995}. There are surely more accurate representations of noise, but this important issue is not our focus. We retain these perturbations for numerical convenience, but stress that the general principles of the TPT framework are independent of any specific form of stochasticity. In the forthcoming experiments, we will refer only to $\sigma_3$ with the understanding that $\sigma_1$ and $\sigma_2$ are adjusted proportionally. 
The discretization we use has strengths and limitations. Given the matrix $L\st$ on this grid, the discretized generator $L$ is just the transpose. To ensure certain properties of solutions, such as positivity of probabilities, $L$ should ideally retain characteristics of the infinitesimal generator of a discrete-space, continuous-time Markov process: rows that sum to zero, and nonnegative off-diagonal entries. Such a discretization is called ``realizable'' \citep{bou-rabee_vanden-eijnden}. One can check that our discretization always satisfies the former property, and so is realizable provided small enough grid spacing $(\delta X,\delta Y,\delta U)$. In our current example, the spacing is not nearly small enough to guarantee this (matrix entries were just as often negative as positive), but results are still accurate, as verified by stochastic simulations to be described in the results section. While we could have used one-sided finite differences to enforce positivity, this would have degraded the overall numerical accuracy of the solutions. We opted instead to zero out negatives, which were always negligible in magnitude.

The discretized Kolmogorov backward equation is $L\qp=0$, augmented with appropriate boundary conditions. The definition of $A$ and $B$ is a design choice that should satisfy three conditions: (1) they are disjoint, (2) $A$ contains the radiative fixed point and $B$ the fixed point of the vacillating regime, and (3) both sets are relatively stable in the chosen noise range. We choose $A$ and $B$ to be ellipses with orientations determined by the covariance of the equilibrium density of the linearized stochastic dynamics about their respective fixed points, as described in the supplement. The choice of the sizes of $A$ and $B$ is a subjective decision which alters the very definition of a reactive trajectory; hence, different sizes emphasize different features of the transition path ensemble, especially in oscillatory systems like this one. We made $A$ and $B$ large enough to enclose the many loops that often accompany the escape from $A$ and the descent into $B$, so that we can focus on the relatively rare crossing of phase space. More sophisticated techniques exist for shrinking the two sets while erasing resulting loops \citep{Lu2014,Banisch2016}; for simplicity, we forgo these techniques for the current study.

Careful discretization is important for constructing the dominant pathways discussed above, i.e. the streamlines $z_t$ satisfying $\dot z_t=J_{AB}$. Standard integration techniques such as Euler or Runge-Kutta will accumulate errors, not only from Taylor expansion but also from the discretized solution of $\qp$, $\qm$ and $\pi$. These can be severe enough to prevent $z_t$ from reaching set $B$. To guarantee that full transitions are extracted, we instead solve shortest-path algorithms on the graph induced by the discretization, as described in \cite{tpt_mjp}. The supplement contains more details on this computation.

\section{Results}

We begin this section by describing the kinematic path characteristics of the process in its three-dimensional phase space, according to the quantities described above. Following this purely geometrical description, we will suggest some dynamical interpretations and compare with previous studies. Finally, we will map statistical features as functions of background parameters.

\begin{figure}
    \caption{The committor gives the probabilistic forecast of the system, and this plot is an empirical demonstration of its predictive capacity. Here, noise is fixed at $\sigma_3=0.5\,m/s/day^{1/2}$, while topographic forcing $h=25\,m$ in (a), (b) and $30\,m$ in (c), (d). The left column shows the forward committor $\qp$ solved by the finite volume method, averaged in the $Y$ direction. The ellipses labeled $A$ and $B$ are projections of the actual sets onto the $XU$ plane, where $X$ and $U$ are the real part of the streamfunction and the mean zonal wind amplitude. Committor values range from 0 (blue) to 1 (red), with the white contour showing the surface $\qp=\frac12$. The right column compares the PDE solution of the committor with a Monte Carlo solution from running many trajectories. For 50 randomly chosen grid points (sampled uniformly across the committor range $(0,1)$), we launched 60 independent stochastic trajectories and counted the fraction that reached set $B$ first. We call this the empirical committor, $\wt\qp$. Plots (b) and (d) show $\wt\qp$ vs. $\qp$ for these 50 random grid cells. The middle red line is the curve $\wt\qp=\qp$, and the envelope around it is the 95\% confidence interval for sampling errors, based on a Gaussian approximation to the binomial distribution.}
    \includegraphics[width=\linewidth]{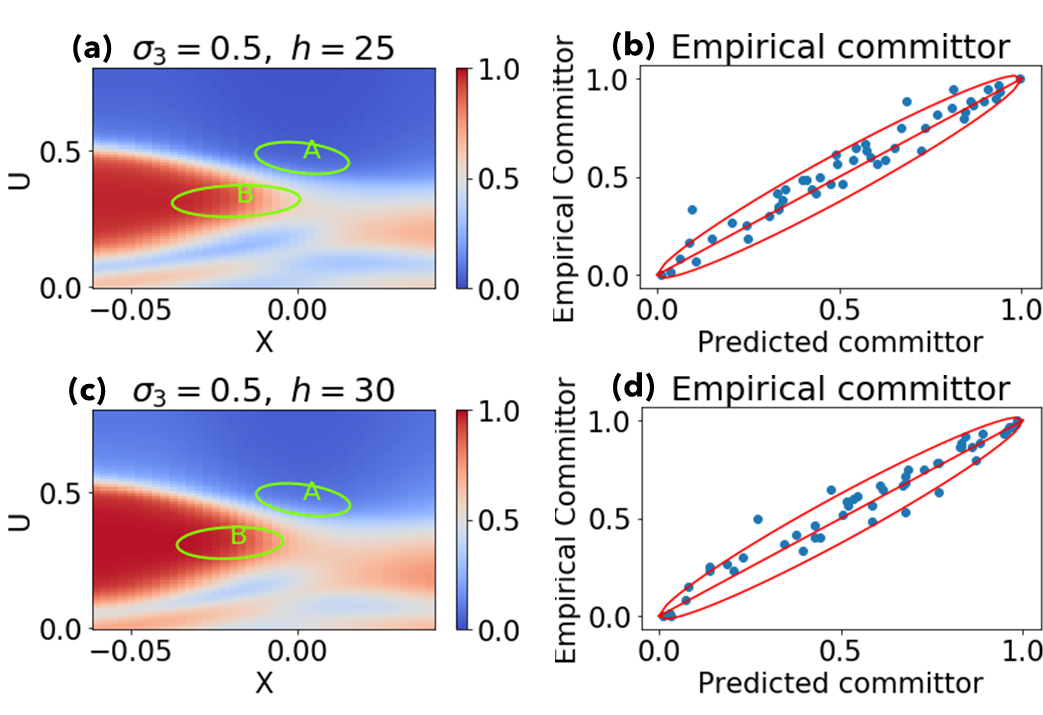}
    \label{fig:q_mcq}
\end{figure}

The Ruzmaikin model is attractive for demonstrating use of the tools introduced in Section 3 due to its low-dimensional state space, in which PDEs can be solved numerically using standard methods such as our finite volume scheme. We tested the committor's accuracy empirically by randomly selecting 50 cells in our grid (this is $0.04\%$ of the grid) and evolving $n=60$ stochastic trajectories forward in time from each, stopping when they reach either set $A$ or set $B$. The fraction of trajectories starting from $z$ that first reach $B$ is taken as the empirical committor at point $z$ and is denoted $\wt\qp$, which we compare with the predicted committor $\qp$ from the finite volume scheme. Figure \ref{fig:q_mcq} clearly demonstrates the usefulness of the committor for probabilistic forecasting. The left column displays the committor calculated from finite volumes, averaged in the $Y$ direction, for two different forcing levels $h$.  The right column shows a scatterplot of $\wt\qp$ vs. $\qp$ at the 50 randomly selected grid cells. We expect the points to fall along the line $\wt\qp=\qp$ with some spread proportional to the standard deviation of the binomial distribution, $\sqrt{\qp(1-\qp)/n}$. Approximating this as a Gaussian, we have plotted red curves enclosing the 95\% confidence interval, which indeed contains approximately 95\% of the data points. Statistical sampling error can explain the observed level of deviation, although discretization error (from solving the PDE and from time-stepping) may also contribute.

\begin{figure}
    \caption{The committor-1/2 surface is the set of all points in state space where the $\qp=\frac12$, and sets $A$ and $B$ have equal probabilities of being visited next. Here the surface is rendered as a set of points and viewed from various vantage points in state space (the supplement shows a video with rotation). The topographic forcing is fixed to $h=25\,m$ and the noise level to $\sigma_3=0.5\,m/s/day^{1/2}$. The blue and red clusters mark sets $A$ and $B$ respectively, centered around the two stable fixed points. The gray points show the location of the surface $\qp=\frac12$. The most striking feature is the ``spiral staircase" structure in the low-$U$ region of phase space. For any given streamfunction phase, the likelihood of heading toward state $A$ or $B$ depends sensitively on $U$, in an oscillatory manner. Even at very low values of $U$, there are narrow channels which are likely to lead back to set $A$ rather than set $B$. This accounts for the blue regions in the lower part of Figure \ref{fig:q_mcq}. These disappear, however, at higher noise, when set $B$ overtakes the lower half of the picture.}
    \includegraphics[width=\linewidth]{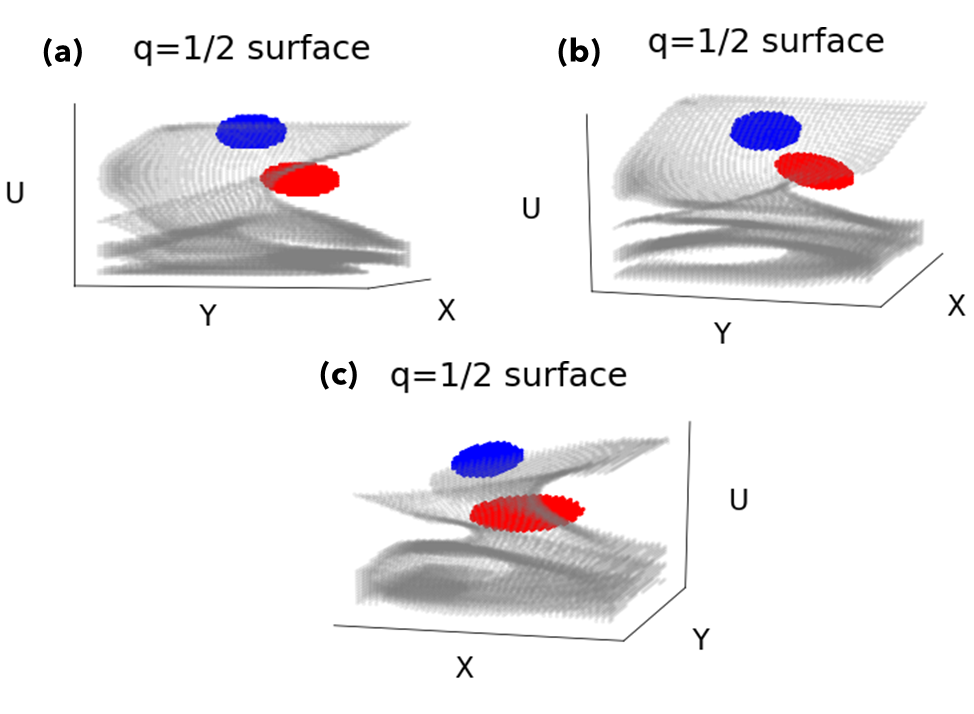}
    \label{fig:isocomm}
\end{figure}

Figure \ref{fig:q_mcq} also shows how $\qp$ responds to increasing $h$, even far below the bifurcation threshold: the committor values throughout state space become rapidly skewed toward unity (meaning more red in the picture). This means that even slight perturbations can kick the system out of state $A$ toward state $B$. Another indicator is the ``isocommittor surface'', the set of points $z$ such that $\qp(z)=1/2$; that is, the system has equal probability of next entering set $A$ or $B$. In the left-hand column this is the set of gray points (averaging out the variable Y). For low forcing, this surface tightly encloses set $B$, meaning the system must wander very close before a transition is imminent. For high forcing values, the isocommittor hugs set $A$ more closely, meaning that small perturbations from this normal state can easily push the system into dangerous territory. In Figure \ref{fig:isocomm}, the isocommittor is shown as a set of gray points in a 3D plot viewed from various vantage points. In the low-$U$ region, the isocommittor resembles a spiral staircase, reflecting the spiral-shaped stable manifold of the fixed point in set $B$. Different initial positions with the same streamfunction phase, differing only slightly in the $U$ direction, can have drastically different final destinations. These spiral surfaces are responsible for the blue lobes in the lower part of Figure \ref{fig:q_mcq}, but they disappear at higher noise. 

\begin{figure}
    \caption{Equilibrium densities $\pi(z)$ (left column) and reactive densities $\rho_R(z)=\pi(z)\qp(z)\qm(z)$ (right column) for two different forcing levels, $h=25\,m$ (top row) and $h=35\,m$ (bottom row). $\pi(z)$ is the long-term probability density of finding the system at point $z$ at any given time; $\rho_R(z)$ is the same probability, but \emph{conditional} on also being reactive at that time, meaning having last visited $A$ and next destined to visit $B$. Dark color indicates higher density. These densities are summed in the $Y$ direction to give a marginal density as a function of $X$ and $U$. The red and blue ellipses show the projected boundaries of sets $A$ and $B$, respectively. These reactive densities capture the patterns of transition path samples shown in Figure \ref{fig:path_samp}, but through continuous fields instead. At low forcing levels, most of the equilibrium density is concentrated around set $A$, whereas higher forcing shifts some of the mass to set $B$. Meanwhile, the characteristic curved shape of transition paths in phase space is borne out by the reactive densities, with a sickle-shaped high-density region bridging the gap between set $A$ and $B$.}
    \includegraphics[width=\linewidth]{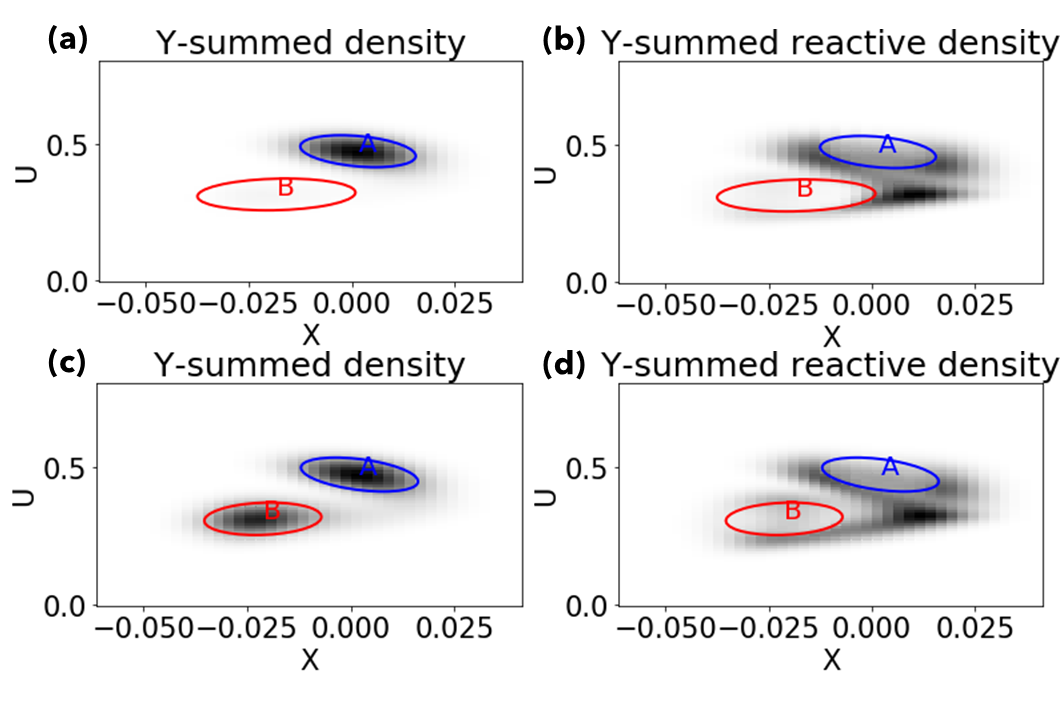}
    \label{fig:xu_densities}
\end{figure}

\begin{figure}
    \caption{Same as Figure \ref{fig:xu_densities}, but projected onto the $XY$ plane, which is the complex plane that characterizes the perturbation streamfunction. The center of set $A$, approximately at $X=Y=0$, corresponds to a zonally symmetric streamfunction with no perturbation. Counterclockwise motion of trajectories around $A$ represents an eastward phase velocity of the streamfunction, which is the dominant modality in the radiative regime. The region of high density above set $A$ shows the phase in the streamfunction at which zonal wind is most likely to begin to weaken. }
    \includegraphics[width=\linewidth]{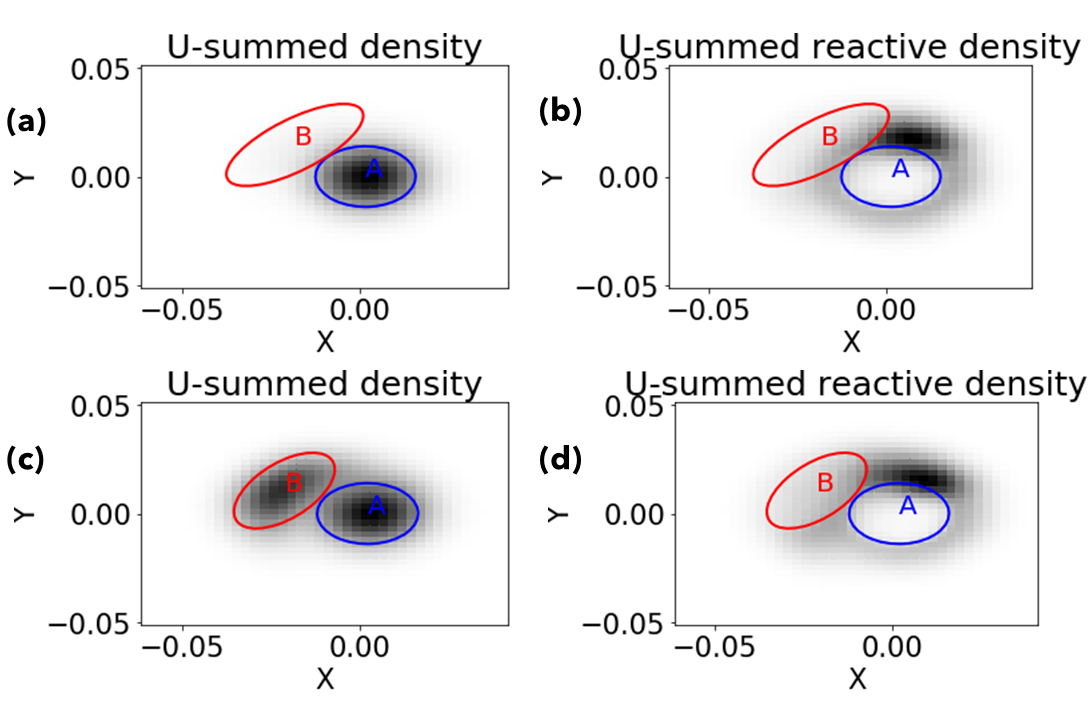}
    \label{fig:xy_densities}
\end{figure}

Figures \ref{fig:xu_densities} and \ref{fig:xy_densities} display numerical solutions of the equilibrium density $\pi$ and reactive density $\rho_R\propto\pi\qm\qp$ for two forcing levels. While $\pi$ indicates where $Z_t$ tends to reside, $\rho_R$ indicates where $Z_t$ resides \emph{given} a transition from $A$ to $B$ is underway. As $h$ increases, even far below the bifurcation threshold, $\pi$ responds strongly, shifting weight toward state $B$. On the other hand, the reactive density displays similar characteristics for all $h$ values. In the $XU$ plane, the two lobes of high reactive density surrounding $A$ indicate that zonal wind tends to remain strong for a while before dipping into the weaker regime. Viewing the same field in the $XY$ plane (Figure \ref{fig:xy_densities}) reveals a halo of intermediate density about set $A$. While many different motions would be consistent with this pattern, the coming figures verify that the early stages of transition have circular loops in the $XY$ plane, meaning zonal movement of the streamfunction's peaks and troughs. The exact streamfunction phase corresponding to the $(X,Y)$ position is calculated as follows. Recall the streamfunction is $\psi'=\re\{\Psi(t)e^{ikx}\cos(\ell y)\}$, where $\Psi=X+iY$. In polar coordinates, $\Psi=\sqrt{X^2+Y^2}e^{i\phi}$, where $\phi=\tan\inv(Y/X)$. The full streamfunction is
\aln{
    \psi'\propto\re\{\Psi e^{ikx}\}&=\re\{\sqrt{X^2+Y^2}e^{i(\phi+kx)}\}\\
    &=\sqrt{X^2+Y^2}\cos(\phi+2\lambda)
}
where $\lambda$ is longitude. 

\begin{figure}
    \caption{Paths of maximal current superimposed on transition path samples. All four Figures shown the same path, but from different vantage points. While the reactive probability density (Figures \ref{fig:xu_densities} and \ref{fig:xy_densities}) says where transition paths spend their time, the reactive current is a vector field of the transition paths' local average directionality. The path shown in a color gradient from blue to red is a streamline of this vector field, representing an ``average" transition path. The path is colored blue where the local committor is less than 0.5, and red otherwise. Note that the path can cross back and forth. The transition from red to blue, where the path first crosses the threshold $\qp=\frac12$ and enters the probabilistic $B$ basin, is marked by a sudden drop in the $U$ variable -- a deceleration in zonal wind. At the same time, the path's rotations about $A$ reverse direction, from clockwise to anti-clockwise, corresponding to a reversal in phase velocity of the streamfunction. This path accurately captures geometric tendencies of actual transition paths; five random samples of reactive trajectories are superimposed in green, the bulk of which cluster around the maximum current path. Panels (a) and (b) show cross-sections in $XU$ and $XY$ space respectively, while (c) shows a three-dimensional view. The parameters are $h=30\,m$ and $\sigma_3=0.5\,m/s/day^{1/2}$. Figure \ref{fig:streamfunctions} shows the corresponding spacetime diagrams of the streamfunction.}
    \centering
    \includegraphics[width=\linewidth]{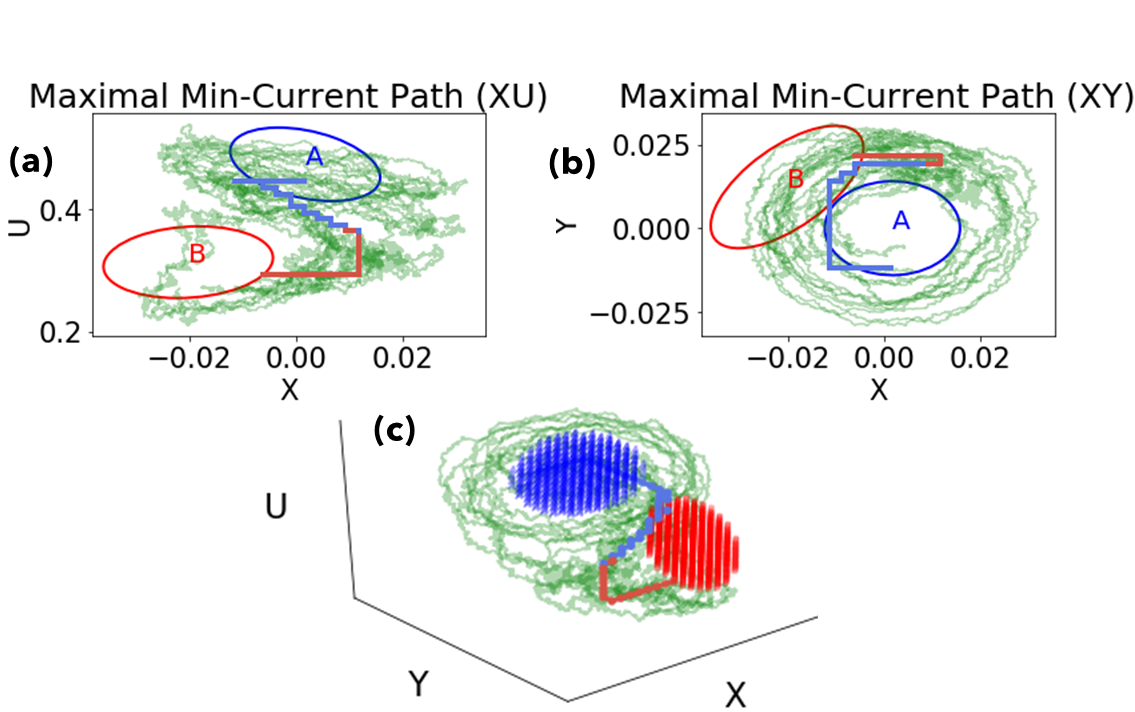}
    \label{fig:dominant_trajectories}
\end{figure}

\begin{figure}
    \caption{Streamfunctions over time corresponding to the trajectories shown in Figure \ref{fig:dominant_trajectories}. As $\psi'\propto(X\cos kx-Y\sin kx)$, the $X$ and $Y$ variables represent the phase of the streamfunction, whose movement we plot over time as a space-time diagram. Panel (a) shows the dominant transition path. The phase velocity is initially eastward, matching with the clockwise rotations in the $XY$ plane as shown in Figure \ref{fig:dominant_trajectories}. The waves then slow down and reverse direction, matching with the anti-clockwise turn and zonal wind drop in Figure \ref{fig:dominant_trajectories}. The vertical axis plays the role of time, but the dominant path technically conveys only geometrical information. Hence, we measure it in discrete steps. Panels (b)-(d) show the streamfunctions over time corresponding to four of the green transition path samples in Figure \ref{fig:dominant_trajectories}, chosen randomly. Most exhibit the same slow-down and reversal behavior exemplified by the dominant path. The exception is sample (b), which turns to the east at the end of its path. This corresponds to the stray green trajectory visible in Figure \ref{fig:dominant_trajectories}(a), which enters set $B$ from above and in the clockwise direction. Samples 1, 2 and 4 undergo some winding before the slowdown, but do slow down every time they reach the same phase.}
    \centering
    \includegraphics[width=\linewidth]{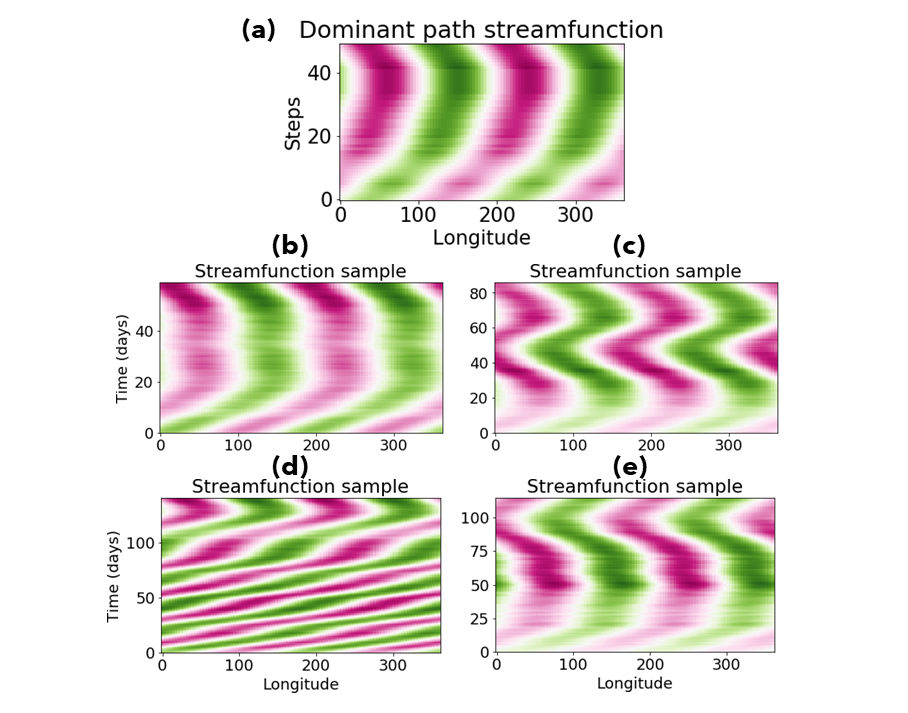}
    \label{fig:streamfunctions}
\end{figure}

The angle from the origin in the $XY$ plane indicates the zonal streamfunction phase, and circular motion indicates zonal movement. (This ``looping" motion is indeed shared by the transition path samples shown in Figures \ref{fig:dominant_trajectories} and \ref{fig:streamfunctions}, to be described later.) 

The darkest (most-trafficked) region of this loop is the sector $\frac{\pi}{4}\lesssim\phi\lesssim\frac{\pi}{2}$.
The relationship between $(X,Y)$ and $\lambda$ indicates $\psi'$ is maximized at longitudes $\lambda=\{-\frac{\phi}{2},\pi-\frac{\phi}{2}\}$. As the maximum reactive density occurs around $\phi=\frac{3\pi}{8}$, the streamfunction peaks are at $\{-\frac{3\pi}{16},\frac{13\pi}{16}\}\approx\{326^\circ,146^\circ\}$. What is the significance of this phase relative to the lower boundary forcing? Recalling the forcing form $\re\{\Psi(z_B,t)e^{ikx}\}=h\,\re\{e^{i2\lambda}\}\propto\cos(2\lambda)$, the bottom peaks are located at $\lambda=\{0,\pi\}$. Hence, the bulk of the transition process happens when the perturbation streamfunction at the mid-stratosphere lags the lower boundary condition by $\frac3{16}\pm\frac1{16}$ of a wavelength. Meanwhile, the $XU$ plane reveals what happens to the zonal wind speed during the SSW transition. The high-reactive density region discussed above coincides with the crescent-shaped bridge of high density between the sets in Figure \ref{fig:xu_densities}. This suggests that in an SSW, the zonal wind weakens while the streamfunction stays in that particular phase window. 

The pictures of reactive density suggest that reactive trajectories tend to loop around set $A$, physically meaning the streamfunction tends to travel in one direction before slowing down, but they technically convey no \emph{directional} information to explicitly support this claim. For this, we turn to the reactive current. We computed the discrete-space effective current matrix $f_{ij}^+$, directly from the finite volume discretization and numerical solutions of $\pi$, $\qp$ and $\qm$. Physically, this matrix represents the flux of a vector field from grid cell $i$ to cell $j$. From this we calculated the maximum-current paths as described in \cite{tpt_mjp} and displayed the results in Figure \ref{fig:dominant_trajectories} for a forcing level of $h=30\,m$ (other levels are qualitatively similar). Both the $XU$ and $XY$ views are shown. Superimposed on these paths are seven actual reactive trajectories that occurred during a long stochastic simulation, to demonstrate features that are captured by the dominant pathways. The dominant path from $A$ to $B$ indeed contains a half-loop in the $XY$ plane in the clockwise direction, which means an eastward phase velocity. With a smaller set $A$, this dominant path would contain more of these loops. However, during the next transition stage, the streamfunction slows to a halt at the phase angle $\phi=\frac\pi2$, doubles back and travels westward as zonal wind loses strength. The smear of high density in the neighborhood $\phi\sim\frac{3\pi}{16}$ therefore comprises not only a precipitous drop in zonal wind (which happens at the edge of that region) but also a backtrack, this time with weaker background zonal wind. This behavior is borne out by the trajectory samples, which vacillate in the upper-middle section of the $XY$ plot. These paths are displayed as space-time diagrams of the streamfunction in Figure \ref{fig:streamfunctions}. In panel (a), the dominant path's two loops correspond to two troughs moving east past a fixed longitude before the slowdown. The random streamfunction trajectories shown in (b)-(g) do not follow this representative history exactly, but they do combine elements of it: steady eastward wave propagation followed by slowdown and reversal. Each stage can have multiple false starts. Notably, the slowdown consistently happens at the same phase, with peaks at $\sim120^\circ E$ and $300^\circ E$, at roughly the same phase as found from the density plots. In fact, the figures show a brief slowdown every time the streamfunction passes this phase. This can be thought of as a representation of blocking events that often accompany sudden stratospheric warmings. The third transition path shown is an exception to the general pattern, making a final turn toward the east instead of to the west. This outlier of a reactive trajectory can also be seen in Figure \ref{fig:dominant_trajectories}, as the single green trajecory that decreases in $X$ before decreasing in $U$ instead of the other way around. 

This kinematic sequence of events has a dynamical interpretation with precedent in prior literature. A critical ingredient of SSW is meridional eddy heat flux, which in this model takes the form $\ov{v'\Phi_z'}\propto hY$ (see the supplement for a detailed derivation.) This term shows up explicitly as a negative forcing on $U$ in equation (15), showing that a reduced equator-to-pole temperature gradient in turn weakens the vortex via the thermal wind relation. The association of heat flux with SSW has been demonstrated in reanalysis \citep{Sjoberg2012} and in detailed numerical simulations of internal stratospheric dynamics, even with time-independent lower boundary forcing \citep{Scott2006}. This relationship favors the phase $\phi=\frac\pi2$ as the most susceptible state for SSW onset, which is exactly picked out by the dominant transition pathway in Figure \ref{fig:dominant_trajectories}.

However, immediately after the wind starts weakening at $\phi=\frac\pi2$, where the streamfunction lines up with its lower boundary condition, the phase velocity reverses, giving rise to the westward lag of $\frac{3\pi}{16}$ we observed in the reactive density. A similar phase lag has also been observed in more detailed numerical studies. For example, \cite{Scott2006} observed a lag of $\frac\pi2$ across the whole stratosphere ($\frac\pi4$ at the mid-level), quite similar to our result. They found that vortex breakup was preceded by a long, slow build-up phase in which the vortex became increasingly vertically coherent, only to be ripped apart by an upward- and west-propagating wave. In an experiment with slowly-increasing lower boundary forcing, \cite{Dunkerton1981} saw a phase lag across the whole stratosphere that increased from $\sim100^\circ$ to $\sim180^\circ$ ($50^\circ$ to $90^\circ$ between the lower boundary and the mid-stratosphere) over the course of the warming event. They attribute this phase tilt to the zonal wind rapidly reversing and carrying the streamfunction along. The weakening zonal wind simply removes the Doppler shift from the Rossby wave dispersion relation, $\omega=Uk-\beta k/(k^2+\ell^2)$, allowing the waves to revert to their preferred westward phase velocity. This balance is also clear in equations (13-14): ignoring the damping and forcing terms, the dynamics read $[\dot X,\dot Y]=[(sU-r)Y,-(sU-r)X]$. For weak $U$, rotation in the $XY$ plane is anti-clockwise and phase speed is westward.


Let us re-emphasize the probabilistic interpretation of reactive density. We have found that transitions from $A$ to $B$ are accompanied by anomalous increases in meridional heat flux. In other words, $\pr\{\text{High heat flux$|$SSW}\}$ is high. As noted earlier, this does not imply that $\pr\{\text{SSW$|$high heat flux}\}$ is also high. The identified streamfunction phase is not a ``danger zone'' in the sense that a trajectory entering this region is at higher risk of falling into state $B$; the committor alone conveys that information. Rather, a trajectory is highly likely to pass through that region \emph{given} that it is reactive. Notably, reactive trajectories are unlikely to take a straight-line path from $A$ to $B$ with $U$, $X$ and $Y$ changing linearly. This unrealistic path would represent a zonally stationary streamfunction growing steadily in magnitude, while zonal wind falls off gradually. At higher noise levels, however, the system would be increasingly dominated by pure Brownian motion, and such a path would become more plausible. 


We now turn to a quantitative comparison of committors and transition rates for different forcing and noise levels. These trends illustrate the effects of modeling choices and global change on the climatology of SSW. Planetary wave forcing, $h$, varies across days and seasons as well as different planets. The strength of additive noise, $\sigma$, is a modeling choice intended to represent gravity wave drag. Different stochastic parametrizations will vary in their effective $\sigma$ value, and it is important to understand the sensitivity of SSW to model choices \citep{sigmond}. Furthermore, long-term climate change may cause both parameters to drift, altering the occurrence of SSW-induced severe weather events.

The measure of the relative ``imminence" of a vacillating solution vs. a radiative solution, as described in the background section on committors, is the equilibrium density-weighted average committor, denoted $\langle\qp\rangle_\pi$. Figure 11(a) shows this quantity for $25\leq h\leq45$ ($m$) and $0.4\leq\sigma_3\leq2.0$ ($m/s/day^{1/2})$. Two trends are clearly expected from the basic physics of the model. First, as seen in Figures \ref{fig:xu_densities} and \ref{fig:xy_densities}, $\langle\qp\rangle_\pi$ should increase with $h$. Second, in the limit of large noise and infinite domain size, the dominance of Brownian motion will smooth out the committor function and make $\langle\qp\rangle_\pi$ tend to an intermediate value between zero and one. On the other hand, as noise approaches zero, the dynamical system becomes increasingly deterministic, and the ultimate destination of a trajectory will depend entirely on which basin of attraction it starts in. The boundary, or \emph{separatrix}, between these two basins is the stable manifold of the third (unstable) fixed point. In the case of a potential system, of the form $\dot x=-\nabla V(x)$, this boundary would be a literal ridge of the function $V$. Our system admits no such potential function, but this is a useful visual analogy. The committor function becomes a step function in the deterministic limit, with the discontinuity located exactly on this boundary. The addition of low noise moves the committor-$\frac{1}{2}$ surface away from the separatrix, possibly asymmetrically: one basin will shrink, becoming more precarious with respect to random perturbations, while the other will expand, becoming a stronger global attractor. Which basin will shrink is not evident \emph{a priori}, so we compute the averaged committor, $\langle\qp\rangle_\pi$, as a summary statistic which will increase when the basin of $B$ expands. 

\begin{figure}
    \caption{Behavior of the committor as a function of forcing $h$ and noise $\sigma_3$. (a) shows the average committor (weighted by equilibrium density): $\int\qp(x)\pi(x)\,dx$ evaluated for a range of $h$ and $\sigma$ values. Panels (b)-(e) show images of the committor for $h=30$ and $\sigma=0.5,\ 0.75,\ 1.0,\ 1.25$. The blue lobes in the lower part of the images, a shadow of the spiral structure from Figure \ref{fig:isocomm}, thin out and disappear with increasing forcing $h$.}
    \includegraphics[width=\linewidth]{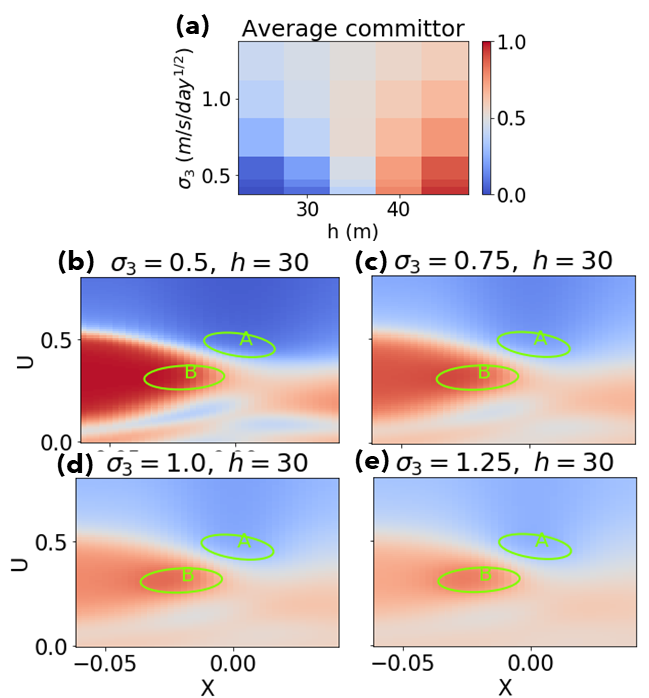}
    \label{fig:q_piavg}
\end{figure}

Figure \ref{fig:q_piavg}(a) plots the trends in $\langle\qp\rangle_\pi$ as a function of $h$ (along the horizontal axis) and $\sigma_3$ (along the vertical axis). The two basic hypotheses are verified: $\langle\qp\rangle_\pi$ increases monotonically as $h$ increases, and $\langle\qp\rangle_\pi\sim\frac12$ as $\sigma$ increases, no matter the value of $h$. The less predictable behavior is in the range $h=35\,m$, $0.75\leq\sigma_3\leq1.0$, where $\langle\qp\rangle_\pi$ displays non-monotonicity with respect to noise, at low noise levels.  As $\sigma_3$ increases, the average committor increases from $\sim0.4$ to $\sim0.6$, and then decreases again. The four committor plots at the bottom of \ref{fig:q_piavg} illustrate the trend graphically. At low noise, the $A$ basin includes winding passageways leading from the small-$U$ region back to $A$. Small additive noise closes them off, effectively expanding the $B$ basin. As noise increases and Brownian motion dominates the dynamics, committor values everywhere relax back to less-extreme values, reflecting the unbiased nature of Brownian motion.

Despite the coarse grid resolution, the first-order effect of noise is clear. At the low and high margins of $h$, where falling into state $A$ and $B$ respectively is virtually certain, an increase in noise decreases this virtual certainty, and the trend continues at larger noise to attenuate $\langle\qp\rangle_\pi$ to its limit of 1/2. The middle $h$ range, however, behaves differently. Whereas $h=35\,m$ appears to balance out the basin sizes at low noise, a slight noise increase tends to kick the system out of the $A$ basin and toward $B$, more so than the other way around.  At higher noise, the committor relaxes back to 1/2. Examining the committor fields, it's clear that the isocommittor surface location does not move back and forth; rather, it moves toward $A$, and then the rest of the field flattens out. 

\begin{figure}
    \caption{Behavior of return times as a function of forcing $h$ and noise $\sigma_3$. Panel (a) shows the average period between the start of one transition event and the start of the next. Red here means many transitions per unit time, both $A\to B$ and $B\to A$. We call this the return time, and calculate it as the reciprocal of $R$, the number of forward (or backward) transitions per unit time. There is clearly a parameter set: $(h,\sigma_3)\approx(35\,m,0.75\,m/s/day^{1/2})$, which optimizes the number of transitions per unit time. Below this noise level, internal variability is scarcely enough to jump between regions. Above this noise level, sets $A$ and $B$ are no longer metastable, and excursions are so wide and frequent that passing from set $A$ to set $B$ is a very spatially restricted event. Panels (b) and (c), below, distinguish forward and backward transition times. Panel (b) shows the expected passage time $T_{AB}$, the interval between the end of a $B\to A$ transition and the end of the next $A\to B$ transition. Panel (c) shows the analogous backward passage time, $T_{BA}$. Note the scales are logarithmic, and here red simply means faster transitions, regardless of which direction is being considered.}
    \centering
    \includegraphics[width=\linewidth]{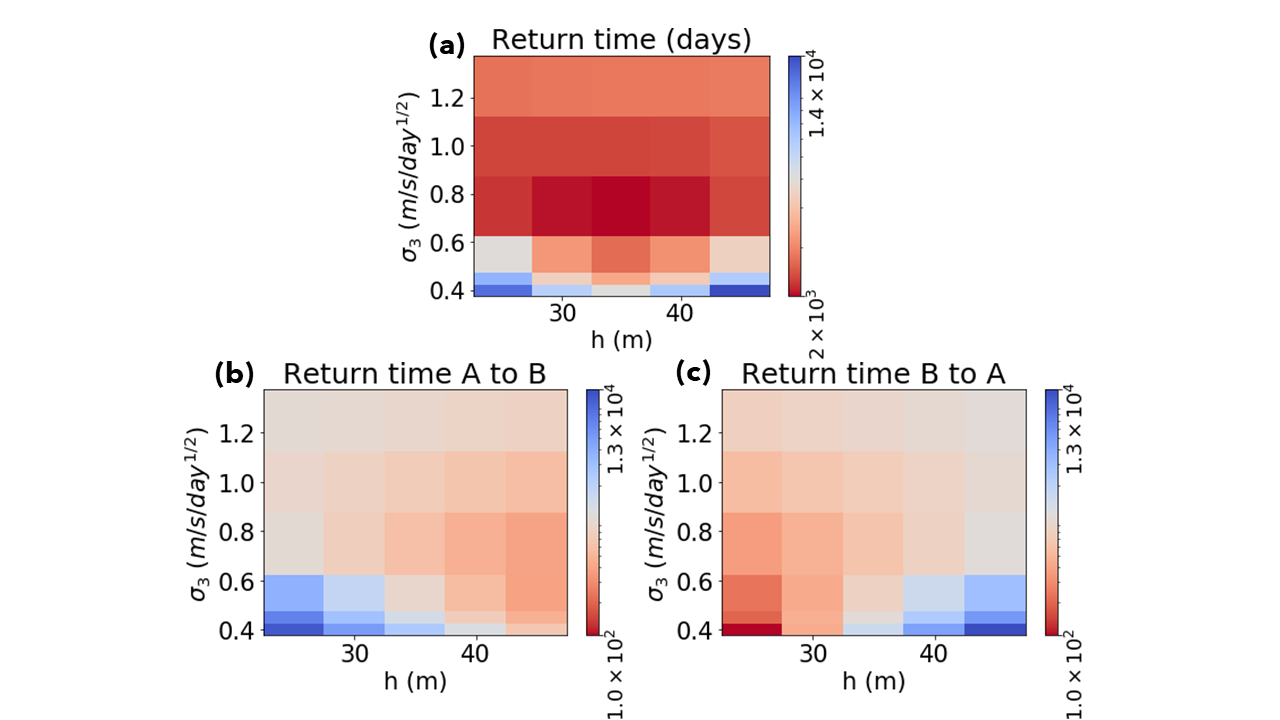}
    \label{fig:rate}
\end{figure}

Figure \ref{fig:rate} shows trends in the return times of SSW with varying $h$ and $\sigma_3$. There are several different return times of interest. The first, shown in panel (a), is the total expected time between one transition event and the next, whose reciprocal is the \emph{rate} $R_{AB}$, the number of transitions per unit time. The return time is a symmetric quantity between $A$ and $B$, since every forward transition is accompanied by a backward one. Among the parameter combinations, $(h,\sigma_3)\approx(35\,m,0.75\,m/s/day^{1/2})$ is the one which minimizes return time, or equivalently maximizes the transitions per unit time. $h=35\,m$ is a forcing level which approximately balances out the time spent between the two sets, making transitions relatively common. At lower noise, transitions are exceedingly rare, and at higher noise the two states cease to be long-lived. However, this symmetric quantity does not capture information about the relative speed of transition from $A$ to $B$ vs. from $B$ back to $A$. Panel (b) shows a different passage time, which is the average time between the end of a backward $(B\to A)$ transition and the end of the next forward ($A\to B$) transition, which we call $T_{AB}$. This is computed as the reciprocal of the rate constant $k_{AB}$, as described in the previous section. In other words, the stopwatch begins when the system returns to $A$ after having last visited $B$, and ends when the system next hits $B$. This metric is asymmetric: a smaller $A\to B$ return time indicates that the forward transition is faster than the backward transition. Panel (b) shows the complementary $B\to A$ return time, $T_{BA}$. Unsurprisingly, an increase in $h$ causes a decrease in $T_{AB}$ and an increase in $T_{BA}$ regardless of the noise level. The noise level has a less obvious effect. Whereas $T_{BA}$ decreases monotonically with increasing noise, regardless of $h$, the forward time $T_{AB}$ is minimized by a mid-range noise level of $\sigma_3\approx0.75\,m/s/day^{1/2}$. This is another reflection of the bias toward state $B$ that is effected by adding noise to a very low baseline.  

\section{ Conclusion}
Transition path theory is a framework for describing rare transitions between states. We have described TPT along with a number its key ingredients like the forward and backward committor functions. While TPT has been applied primarily to molecular systems, we believe it offers valuable insight into climate and weather phenomena such as sudden stratospheric warming, primarily through committors, reactive densities, and reactive currents. Of interest apart from its role in TPT, the forward committor defines an optimal probabilistic forecast, borne out by direct numerical simulation experiments. The reactive densities and currents describe the geometric properties of dominant transition mechanisms at low noise. In applying TPT to a noisy, truncated Holton-Mass model, we find that transitions tend to begin with a drop in mean zonal wind and a reversal of the streamfunction's phase velocity at a particular streamfunction phase. This is consistent with the significance of blocking precursors to SSW as found in \cite{martius}, insofar as this idealized model can represent them. We also find that noise has a non-monotonic effect on the overall preference for a vacillating state, measured by the average committor. At a forcing of $h\approx35\,m$, where the isocommittor surface (essentially the basin boundary of the deterministic dynamics) divides the space approximately in half, we find that raising the noise tilts the balance decisively toward the vacillating solution. Still larger noise evens the whole field out. The transition rate constant shows a similar dependence on $h$ and $\sigma_3$. 

In future work, we plan to scale these methods to more realistic and complex models as well as observational data, where predicting SSW remains an active area of research. In this high-dimensional setting, the generator may be unknown or computationally intractable. Any state space with more than $\sim5$ degrees of freedom is beyond the reach of a finite-volume discretization, because the number of grid cells increases exponentially with dimension. However, there is a generic insight that physical models evolve on very low-dimensional manifolds within the full available state space. A growing body of research in molecular dynamics \citep{dga}, fluid dynamics \citep{giannakis_kolchinskaya_krasnov_schumacher_2018,Froyland2018}, climate dynamics \citep{nlsa,Sabeerali2017}, and general multiscale systems \citep{Harlim2018,Berry2015,Giannakis2015,Giannakis2019} exploits the intrinsic low-dimensionality to represent the infinitesimal generator more efficiently. While here we represented the generator as a finite-volume or finite-difference operator on a grid, one can also write it in a basis  of globally coherent functions, such as Fourier modes, or more generally harmonic functions on a manifold. Given only data, without an explicit form of the dynamics, this manifold and the basis functions can be estimated from (for example) the diffusion maps algorithm, and the generator's action on this basis can be approximated from short trajectories. These ideas can be applied to computing the dynamical statistics that have been the focus of this paper \citep{dga}. We hope that these techniques will enable more efficient observation strategies for targeted data assimilation procedures with the goal of tracking the progression of specific extreme events, including hurricanes and heat waves as well as sudden stratospheric warming.

%
\datastatement
Data from this study will be made available upon request.

%
\acknowledgments
Justin Finkel was funded by the Department of Energy Computational Science Graduate Fellowship under grant DE-FG02-97ER25308. We acknowledge support from the National Science Foundation under NSF award number 1623064. Jonathan Weare was supported by the Advanced Scientific Computing Research Program within the DOE Office of Science through award DE-SC0020427. The reviewers of the paper provided invaluable insights on physical interpretation of the mathematical results. In particular, they helped us to recognize the connection between our computed maximum-flux reactive pathway and anomalous heat fluxes which weaken the zonal wind and reverse the streamfunction's phase velocity. We thank Cristina Cadavid, Thomas Birner and Paul Williams for helpful clarifications about previous work. We thank Eric Vanden-Eijnden, Mary Silber, Robert Webber, Erik Thiede, Aaron Dinner, Tiffany Shaw, and Noboru Nakamura for useful discussions throughout this project.  The University of Chicago's Research Computing Center provided the computational resources to greatly expedite the calculations done here.

%
\appendix
Below are the numerical coefficients used in the reduced-order Ruzmaikin model, with very similar values to \cite{ruz} and \cite{birner_williams}. The relationship with physical parameters is described in the appendix of \cite{ruz}. Note that our notation differs slightly: following \cite{birner_williams}, we write the topographic forcing in terms of $h$ rather than $\Psi_0=\frac{gh}{f_0}$, a difference that results in numerical factors of $\sim1000$ depending on the convention used. 
\alnn{
\tau_1&=122.6\\
\tau_2&=30.4\\
r&=0.63\\
s&=1.96\\
\xi&=1.75\\
\delta_w&=70.84\\
\zeta&=240.54\\
U_R&=0.47\\
\eta&=9.13\times10^4\\
\delta_\Lambda&=4.91\times10^{-3}\\
\dot\Lambda&=0\\
}






%
%
%
\bibliographystyle{ametsoc2014}
\bibliography{references}

%

%

\end{document}